# Calculation and evaluation of cross-sections for p+$^{184}$W reactions up to 200MeV [*]

SUN Jian-Ping(孙建平)[1] ZHANG Zheng-Jun(张正军)[1;1)] HAN Yin-Lu(韩银录)[2]

[1] Department of Physics, Northwest University, Xi'an 710069, China

[2] Science and Technology on Nuclear Data Laboratory, China Institute of Atomic Energy, Beijing 102413, China

1) E-mail: Zhangzj@nwu.edu.cn

**Abstract:** The cross-sections of proton-induced reactions on $^{184}$W at incident proton energy below 200MeV are calculated and analyzed including reaction cross-sections, elastic scattering angular distributions, energy spectra and double differential cross section. Nuclear theoretical models which integrate the optical model, distorted born wave approximation theory, the intra-nuclear cascade model, the exciton model, the Hauser-Feshbach theory and the evaporation model are used in the reactions. Theoretical results are compared with the existent experimental data.

[*] Supported by National Basic Research Program of China, Technology Research of Accelerator Driven Sub-critical System for Nuclear Waste Transmutation (No. 2007CB209903) and Strategic Priority Research Program of the Chinese Academy of Sciences, Thorium Molten Salt Reactor Nuclear Energy System (No.XDA02010100).

# 1 Introduction

Nuclear power is a kind of efficient and clean energy. The accelerator driven clean nuclear power system (ADS) is a reasonable choice to solve the final processing of long-lived nuclear waste and the effective use of the resources, and will become the transition between current commercial nuclear energy system and fusion energy system. The abilities of nuclear waste transmutation and fuel deepened utilization of ADS system largely depend on the number and energy distribution of external neutron sources maintain the subcritical reactor of ADS and are generated from the high energy proton bombardment of heavy nuclei. Micro data are basis for the whole ADS system in the high energy proton nuclear reactions. So studying theory and micro data of medial or high energy protons and heavy metals reaction is important.

Tungsten (W) plays an important role in the target materials for neutron source, and the proton-induced reactions data on W in the energy range from the threshold energy to 200MeV are necessary for ADS, particularly the data of proton-induced energy-angle correlated spectra and double differential cross-sections of particles (neutron, proton, deuteron, triton, helium-3 and alpha-particle) emission.

The work mainly uses optical model, exciton model, evaporation model and intra nuclear cascade to describe the different stages of the nuclear reaction. Overall consideration, the theories and methods further are improved in intermediate and high energy reaction, and then the micro data can be used for ADS target and other device design are obtained from the theoretical calculations. In this work, the proton-induced cross-sections, energy spectra and double differential cross sections are calculated and analyzed of p+ $^{184}$W reactions [1].

The theoretical methods and model parameters are shown in Section 2. The comparisons and analysis between the calculated results and the experimental data are given in Section 3. Finally, Section 4 gives the conclusion.

# 2 Theoretical models and model parameters

The optical model [2, 3] is used to describe the experimental data of proton-induced total, non-elastic, elastic cross sections and elastic-scattering angular distributions, and to calculate the transmission coefficients for the compound nucleus and the pre-equilibrium emission process. [4]

The optical potential expression is

$$V = V_r + i(W_s + W_v) + U_{so} + V_c, \qquad (1)$$

where $V_r$ is the real part potential, $W_s$ and $W_v$ are the imaginary part potentials of surface absorption and volume absorption, $U_{so}$ is the spin-orbit couple potential and $V_c$ is the Coulomb potential.

The real part of optical model potential is

$$V_r = -\frac{V_r(E)}{1+\exp[(r-R_r)/a_r]}, \qquad (2)$$

The imaginary part of surface absorption is

$$W_s = -4W_s(E)\frac{\exp[(r-R_s)/a_s]}{\{1+\exp[(r-R_s)/a_s]\}^2}, \qquad (3)$$

The imaginary part of volume absorption is

$$W_v = -\frac{W_v(E)}{1+\exp[(r-R_v)/a_v]}, \qquad (4)$$

The spin-orbit potential is

$$U_{so} = -\frac{2(V_{so}+iW_{so})}{a_{so}r}\frac{\exp[(r-R_{so})/a_{so}]}{\{1+\exp[(r-R_{so})/a_{so}]\}^2}[j(j+1)-l(l+1)-s(s+1)], \qquad (5)$$

The Coulomb potential is

$$V_C = \begin{cases} \dfrac{zZe^2}{r} & \text{if } r \geq R_C \\ \dfrac{zZe^2}{2R_C}(3-\dfrac{r^2}{R_C^2}) & \text{if } r < R_C \end{cases}, \qquad (6)$$

$V_r(E)$ is the real part of potential depth. Its expression is

$$V_r(E) = V_0 + V_1 E + V_2 E^2 + V_3(N-Z)/A + V_4 Z/A^{1/3}, \qquad (7)$$

$W_s(E)$ and $W_v(E)$ represent potential depths of the imaginary parts of surface and volume absorption. They are

$$W_s(E) = \max\{0, W_{s0} + W_{s1}E + W_{s2}(N-Z)/A\}, \qquad (8)$$

$$W_v(E) = \max\{0, W_{v0} + W_{v1}E + W_{v2}E^2\}, \qquad (9)$$

where $N$, $Z$ and $A$ are neutron, charge and mass numbers of target, respectively. $a_r$ in Eq.(2) and $a_{so}$ in Eq.(5) are the diffusive widths of real part and the spin-orbit couple potential separately. The diffusive widths of surface and volume absorption potential in Esq. (3) and (4) are $a_s$ and $a_v$. $z$ represents the charge number of incident particle. $E$ is proton's incident energy in the center of mass system. The spin-orbit couple potential is $V_{so}$.

The radii are given by

$$R_i = r_i A^{1/3}, i = r,s,v,so,c, \qquad (10)$$

where $r_r, r_s, r_v, r_{so}$ and $r_c$ are the radii of the real part, the surface absorption, the volume absorption, the spin-orbit couple and the Coulomb potential, respectively.

The units of the length parameters $r_r, r_s, r_v, r_{so}, r_c, a_r, a_{so}, a_{s0}, a_{s1}, a_{v0}, a_{v1}$ are in fermi (fm), the potentials $V_r, W_s, W_v, U_{so}, V_c$ are in MeV and the energy $E$ is also in MeV. [5]

The best proton optical model potential parameters are obtained for $^{184}$W in Table 1 by the code APMN [6] to fit the experimental data.

Table 1. The proton optical model potential parameters obtained.

| | | | |
|---|---|---|---|
| $V_0$ | 44.98434 | $r_r$ | 1.26146 |
| $V_1$ | -0.28003 | $r_s$ | 1.06554 |
| $V_2$ | 0.0004 | $r_v$ | 1.88632 |
| $V_3$ | -45.53588 | $r_{so}$ | 1.26146 |
| $V_4$ | 0.13797 | $r_c$ | 1.51263 |
| $W_{s0}$ | 7.71606 | $a_r$ | 0.67259 |
| $W_{s1}$ | 0.02640 | $a_{so}$ | 0.67259 |
| $W_{s2}$ | 11.83579 | $a_v$ | 0.20000 |
| $W_{v0}$ | 0.59395 | $a_s$ | 0.62640 |
| $W_{v1}$ | 0.005150 | $V_{so}$ | 6.2 |
| $W_{v2}$ | -0.00076 | $W_{so}$ | 0.0 |

The code DWUCK4 [7] is used to calculate cross-sections and angular distributions of direct inelastic scattering to low-lying states in the theory of Distorted Wave Born Approximation (DWBA). Direct reaction and multiple collisions are important in the proton-induced reactions at energy above 50MeV. The discrete levels of residual nuclei in the direct non-elastic scattering process considered to calculate are given in the following Table 2. The discrete levels are taken into account from ground ($0^+$) to the sixteenth (1.5702 $2^+$) excited state for $^{184}$W, levels above the highest excited state are assumed to be overlapping and level density formula is used. In the spectrum calculation, the contribution of each level is dealt with the Gauss broadening.

Table 2. Discrete levels in direct non-elastic scattering process.

| Energy (MeV) | $J^\pi$ | Energy (MeV) | $J^\pi$ |
| --- | --- | --- | --- |
| 0.1112 | 2 + | 1.3604 | 4 + |
| 0.3641 | 4 + | 1.3863 | 2 + |
| 0.7483 | 6 + | 1.4250 | 3 + |
| 0.9033 | 2 + | 1.4310 | 2 + |
| 1.0060 | 3 + | 1.4770 | 6 + |
| 1.1214 | 2 + | 1.5233 | 3 + |
| 1.1338 | 4 + | 1.5369 | 4 + |
| 1.2949 | 5 + | 1.5702 | 2 + |

Intra nuclear cascade model mechanism [8] can be regarded as the supplement and amendment for direct reaction mechanism. It is applied to consider in one to four cascade nucleons emission process. While in the first to fifth particle emission processes, the exciton model [9] is used.

In pre equilibrium mechanism, the improved Iwamoto-Harada model [10, 11] is included in the exciton model for the light composite particle emissions. Zhang et al. [10, 12] improved Iwamoto et al. pick-up mechanism to reduce the pre-formation probabilities. Shen [13] gave the results of a composite particle projectile considering pick-up-type reactions with one and two particles above the Fermi sea by energy-averaged and energy-angle correlated kernels, respectively. This improved model takes into account the possibility that some of particles that form the complex particles may come from levels below the Fermi level. This model is used to improve the energy spectra and the double differential cross-sections in the reactions in compound nucleus.

Considering restriction of Pauli principle, after amendment, the density at exciton state (p, h) of compound nuclear system (Z, A, U) eventually is expressed by following formula

$$\omega(Z,A,U,p,h) = \frac{g(gU_e)^{h'}[gU_e - A(p,h)]^{p'-1}}{p!h!(p+h-1)!} f(U_e), \tag{11}$$

When the value of $gU_e$ is less than or equal to $A(p,h)$, $\omega(Z,A,U,p,h)=0$. The correlation of g with excitation energy is studied and the new single level density formula is proposed in the work.

$$g(U) = \frac{g_0(1-e^{-U_e \times ccg1})}{U_e \times ccg1}, U_e = U - \Delta, \tag{12}$$

where $g_0$ is equal to A/13, U is excitation energy, delta is correction of energy and *a* is parameter of level energy density.

$$f(U_e) = 0.06\pi^2/(aU_e)^{1/8}, \tag{13}$$

$$A(p,h) = \frac{1}{2}p(p-1) + \frac{1}{2}h(h-1), \tag{14}$$

$$h' = \min(p,h), p' = \max(p,h), \tag{15}$$

Through calculation actually, parameter ccg1 is determined 0.05.

The system tends to balance at each intermediate stage, tending to be balanced and emitting particle is competitive. The particle emissions are considered up to 18 times evaporation process. The equilibrium emissions are calculated by using the evaporation model [14] for the first to the eighteenth particle emission and Hauser-Feshbach theory with the width fluctuation correction [15, 16] for the first particle emission in the low-energy region.

The energy spectrum formula can be given as follow

$$\frac{d\sigma}{d\varepsilon_b} = \sum_{J\pi} \sigma_a^{J\pi} \sum_n P^{J\pi}(n) \frac{W_b^{J\pi}(n,E',\varepsilon_b)}{W_T^{J\pi}(n,E')}, \tag{16}$$

where $\sigma_a^{J\pi}$ is the absorption cross-section, $W_T^{J\pi}(n,E')$ is the total emission rate at n-exciton state and $W_b^{J\pi}(n, E', \varepsilon_b)$ is the emission rate of emitted particle b with outgoing energy $\varepsilon_b$, $E'$ is the excitation energy of the pre-equilibrium system. The occupation probability $P^{J\pi}(n)$ of the $n$ exciton state in the $(J, \pi)$ channel is obtained by solving the $J$-dependent exciton master equation to conserve the angular momentum in the pre-equilibrium reaction processes. $W_T^{J\pi}(n, E')$ and $W_b^{J\pi}(n, E', \varepsilon_b)$ of b particle are given as follows

$$W_T^{J\pi}(n, E') = \sum_b \int W_b^{J\pi}(n, E', \varepsilon_b) d\varepsilon_b, \tag{17}$$

$$W_b^{J\pi}(n, E', \varepsilon_b) = \frac{2s_b + 1}{\pi^2 \hbar^3} \mu_b \varepsilon_b \sigma_b^{J\pi}(\varepsilon_b) \sum_{lm} F_{lm}^b(\varepsilon_b) Q_{lm}^b(n) \frac{\omega(p-1, h, E' - \varepsilon_b - B_b)}{\omega(n, E')},$$

$$\tag{18}$$

where $s_b$ and $\mu_b$ are the spin and reduced mass, $\sigma_b^{J\pi}(\varepsilon_b)$ is the inverse cross-section of b particle, $F_{lm}^b(\varepsilon_b)$ stands for the formation factor of the emitted particle b with the configuration $[l, m]$ from the pick-up mechanism of improved Iwamoto-Harade model, and $F_{lm}^b(\varepsilon_b) = 1$ for nucleon, and for light composite particles (deuteron, triton, helium and alpha-particle) [10, 12]. The configuration $[l, m]$ is composed of $l$ particles above Fermi level and $m$ particles blow. $Q_{lm}^b(n)$ is the combination factor to account the nucleon type that composed the cluster for memory by excitation system, $p$ and $h$ are numbers of particles and holes, $n = p + h$ and $\omega(n, E')$ is the exciton state density. The explicit expressions of $F_{lm}^b(\varepsilon_b)$, $Q_{lm}^b(n)$ and $\omega(n, E')$ can be found in Ref. [10].

The double differential cross-sections is provided by the equation

$$\frac{d^2\sigma}{d\Omega d\varepsilon_b} = \frac{1}{4\pi} \left( \frac{d\sigma^{PE}}{d\varepsilon_b} + \frac{d\sigma^{EQ}}{d\varepsilon_b} \right) \frac{a}{\sinh(a)} [\cosh(a\cos\theta) + f_{PE} \cdot \sinh(a\cos\theta)], \tag{19}$$

Where, $\theta$ is the emission angle in the center of mass frame and $a$ is the slope parameter depending on the incident particle type and energy, the target nucleus and the emitting channel. It can be calculated in the procedure in Ref. [17]. The $f_{PE}$ parameter is the fraction of particle emission apart from equilibration process. It can be calculated as follows

$$f_{PE} = \frac{(d\sigma/d\varepsilon_b)_{PE}}{(d\sigma/d\varepsilon_b)_{PE} + (d\sigma/d\varepsilon_b)_{EQ}}, \qquad (20)$$

Where, PE and EQ stand for the pre equilibrium emission and equilibrium emission respectively. The double differential cross-sections can be calculated by the above energy spectra.

All reaction cross sections for proton induced on $^{184}$W at energy up to 500MeV are studied by using the optical model theory. The credible optical model potential parameters for total reaction cross section and main reaction channels are obtained and they are used as input parameters of the code MEND [18] to calculate the direct reaction cross section, pre equilibrium and equilibrium emission. The total cross section is the sum of all reaction cross sections of various mechanisms.

## 3  Theoretical results and analysis

The partial cross-sections, particle emitting spectrums, residual nuclear production cross-sections of proton-induced on $^{184}$W reactions are calculated. The calculated results are compared with the existent experimental data.

The fission cross sections of p+$^{184}$W, $^{208}$Pb, $^{209}$Bi and $^{202}$Hg reactions are calculated using the energy density that adopt Gilbert-Cameron formula and the fission cross section formulas [19-22] and compared with experimental data in Fig. 1. These experimental data of fission cross sections for the proton-induced on $^{208}$Pb and $^{209}$Bi reactions were found out in Ref. [23], while there are no data on $^{184}$W and $^{202}$Hg reactions. In the chart, fission cross section decrease quickly with decreasing of proton number Z and mass number A of target nucleus. The fission cross section for p+ $^{184}$W is only 30mb at the energy value 300MeV and it is fewer than 2 percent of reaction cross section.

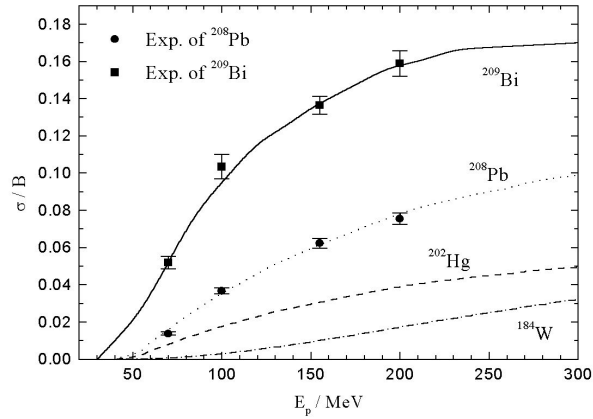

Fig.1. Calculated fission cross sections compared with experimental data (symbols) of p+$^{208}$Pb and $^{209}$Bi reactions. The theoretical results for p+$^{202}$Hg and $^{184}$W reactions are given for lack of experimental data.

The comparison of calculated results of proton reaction cross-sections with experimental data for $^{184}$W is given in Fig. 2. The experimental data of cross sections for p+$^{nat}$W reaction in Ref. [24] are used, since there are no data for $^{184}$W. The calculated results of proton reaction cross-sections for $^{184}$W are in good agreement with the experimental data of natural W. The calculated proton cross-sections are compared with experimental data [24-26] for p+$^{181}$Ta reaction in Fig. 2 and the theoretical results are consistent with these experimental data.

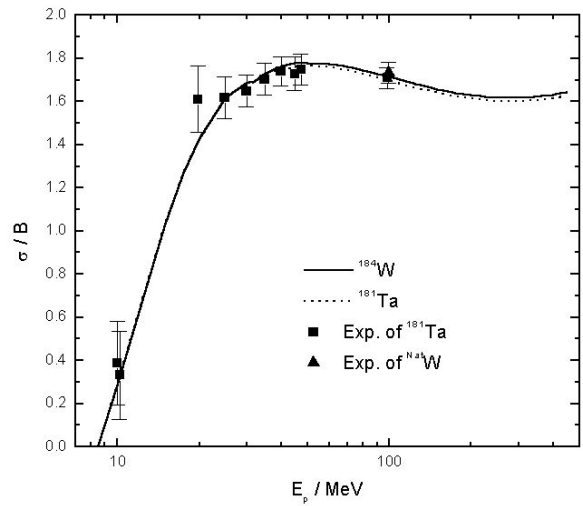

Fig.2. Calculated proton reaction cross-sections (solid line) compared with experimental data (full triangle) for p+$^{184}$W reaction and reaction cross-sections (dotted line) compared with experimental data (full square) for p+$^{181}$Ta reaction.

The comparisons between the calculated results and experimental data of elastic scattering angular distributions for p+$^{184}$W and p+$^{181}$Ta reactions are shown in Fig. 3. The experimental data of p+$^{181}$Ta at the energy 146MeV and 340MeV are from [27, 28] and the data for p+$^{nat}$W reaction at 340MeV are in Ref. [28]. The calculated results are reasonable agreement with experimental data except for the experimental data for $^{184}$W which are larger than the theoretical results at lower degrees.

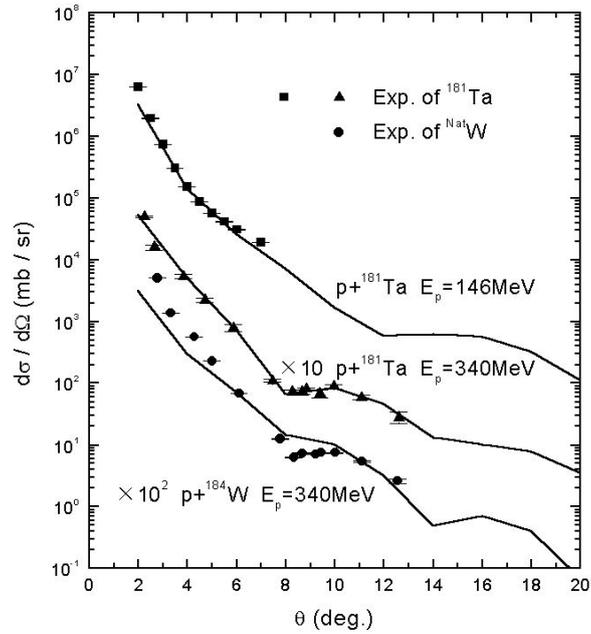

Fig.3. Calculated proton elastic scattering angular distribution (solid line) compared with experimental data (symbols) for p+$^{181}$Ta and p+$^{184}$W reactions at different energies.

The optical model potential parameters of $^{184}$W are proved credible from Fig. 2 and Fig. 3. The optical model potential depth is dependence of mass number A and neutron number N of target nuclei, this set of optical model potential parameters is used in p+$^{184}$W reactions. The results not only reproduced to experimental data, but also displayed some law of nuclear reaction and can be used to predict the reaction without experimental data.

The calculated cross-sections for $^{184}$W (p, kn) reaction channels are shown in Fig. 4. There are no experimental data for $^{184}$W (p, kn) reactions, k is equal to 1,2,3,4,5,6,7,8 and 9. The results calculated for $^{184}$W (p, pkn) reaction cross-sections are given in Fig. 5. At the low proton incident energy, reaction channel (p, 1n) in Fig. 4 open at first that is only one neutron emitted as same as reaction channel (p, p1n) open firstly in Fig. 5. When the energy increases, other channels will open gradually. At the same time, competition exists among these reaction channels. The reaction channel will reduce in a certain level along with a new channel opening.

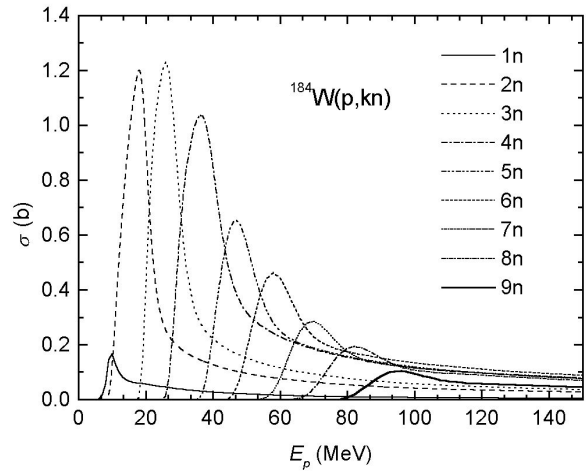

Fig.4. Calculated reaction cross-sections for $^{184}$W (p, kn) channels. k=1, 2, 3, 4, 5, 6, 7, 8, 9.

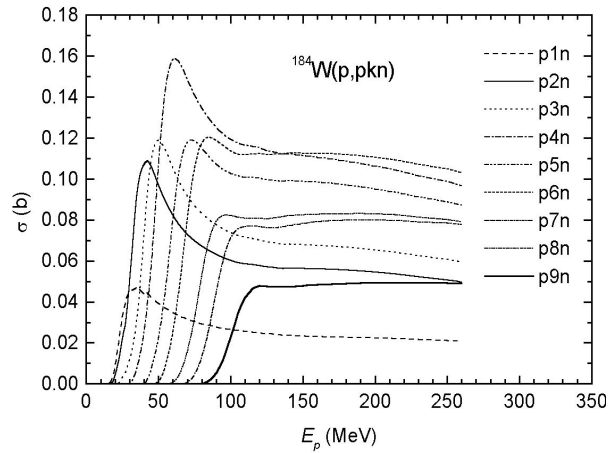

Fig.5. Cross-sections for $^{184}$W (p, pkn) reaction channels are obtained. k=1, 2, 3, 4, 5, 6, 7, 8, 9.

The cross-sections calculated for $^{184}$W(p, pn)$^{183}$W, $^{184}$W(p, 2p3n)$^{180}$Ta, $^{184}$W(p, 2p4n)$^{179}$Ta, $^{184}$W(p, 3p4n)$^{178}$Hf, $^{184}$W(p, 3p8n)$^{174}$Hf, $^{184}$W(p, 3p10n)$^{172}$Hf reactions are shown in Fig. 6. Long-loved radioactive reaction products and production cross sections in the p+$^{184}$W reactions are shown in Table 3. It can be seen that the half-life of these residual nuclei are greater than one year and these cross section values of productions are larger than 1mb in Fig.6. The waste material will have radioactivity for the long-lived radioactive products.

Table 3. Long-lived radioactive products are shown in p+$^{184}$W reaction.

| Residual nuclei | Channels | Half-life (year) |
| --- | --- | --- |
| $^{183}$W | $^{184}$W(p, pn)$^{183}$W | $1.1\times10^{17}$ |
| $^{180}$Ta | $^{184}$W(p, 2p3n)$^{180}$Ta | $1.2\times10^{15}$ |
| $^{179}$Ta | $^{184}$W(p, 2p4n)$^{179}$Ta | 1.82 |
| $^{182}$Hf | $^{184}$W(p, 3p)$^{182}$Hf | $9\times10^{6}$ |
| $^{178}$Hf | $^{184}$W(p, 3p4n)$^{178}$Hf | 31 |
| $^{174}$Hf | $^{184}$W(p, 3p8n)$^{174}$Hf | $2\times10^{15}$ |
| $^{172}$Hf | $^{184}$W(p,3p10n)$^{172}$Hf | 1.87 |

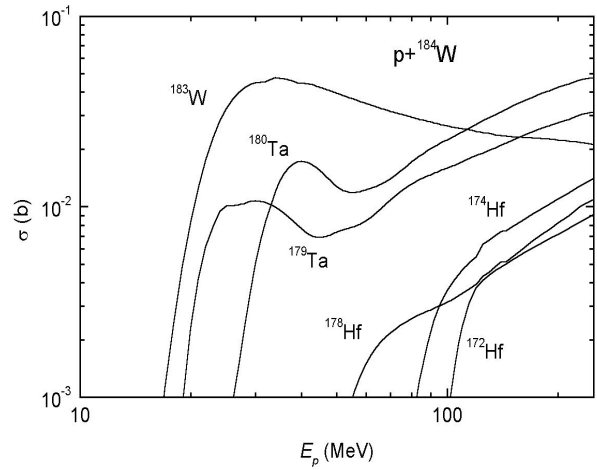

Fig.6. Cross sections of long-lived radioactive reaction products in the p+$^{184}$W reactions are shown.

The energy spectra and double differential cross sections of emission neutron, proton, deuteron, triton, helium and alpha particle for p+$^{184}$W reactions are calculated by theoretical models and codes. The calculated energy spectra of neutron emission for p+$^{184}$W at incident proton energy 50,100,150,200 and 250MeV are shown in Fig. 7. The shape of calculated results curve of energy spectra for neutron emission at the various energies are similar. The calculated results are from the contribution of evaporation model below neutron emission energy 15.0MeV, and are from the contribution of pre-equilibrium process related with the density of states above the energy 15.0MeV.

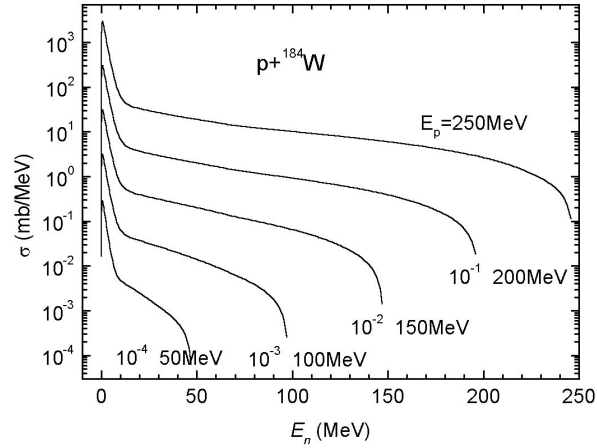

Fig.7. Theoretical energy spectra of neutron emission are calculated in p+$^{184}$W reaction.

The comparison of calculated double differential cross section of neutron emission with experimental data taken from [29] at the incident proton energy 113MeV are given in Fig. 8. The original data are multiplied by 0.66 for the processing problem [30]. There are no experimental data at other energies. From Fig. 8, the calculated results are in good agreement with experimental data at emission angles 7.5, 30.0, 60.0 and 150.0°. There are no experimental data of double differential cross sections of proton, deuteron, triton, helium and alpha emission for p+$^{184}$W reactions.

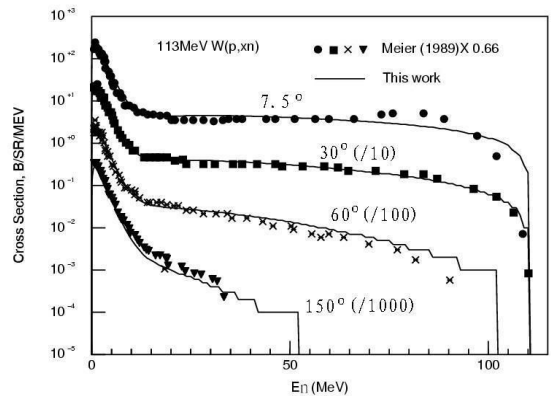

Fig.8. Calculated double differential cross section of neutron emission (solid line) compared with experimental data (symbols) in p+$^{184}$W reaction at the incident energy 113MeV at 7.5, 30.0, 60.0 and 150.0°.

The double differential cross sections of proton emission at the energy 113.0MeV are calculated in Fig. 9. The peaks of the curves at high proton emission energy are contributed from the direct reaction mechanism. From the chart, the spectra at large angles become soft. It indicates that pre equilibrium mechanism contribution is very great at high energy, and the cross sections at small angles will be small.

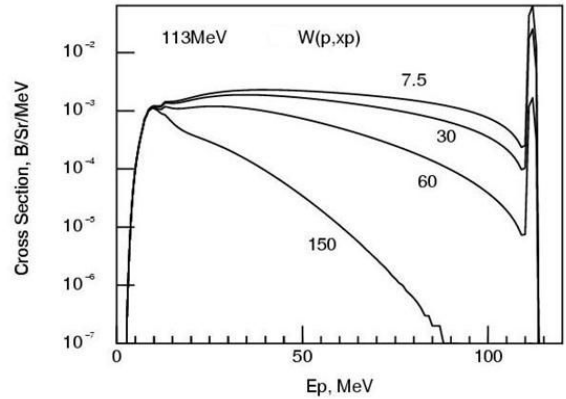

Fig.9. Double differential cross sections of proton emission for p+$^{184}$W reactions are calculated at the incident energy 113.0MeV.

The calculated results of the double differential cross sections of deuteron, triton, helium, alpha particle emissions for p+$^{184}$W reactions are given in Figs. 10-13 at the proton incident energy 113.0MeV. The calculated results are at angles 7.5, 30, 60 and 150º. Since there are no experimental data, the shape of calculated results curve are similar to each other. It is needed to provide more and accurate experimental data in the future.

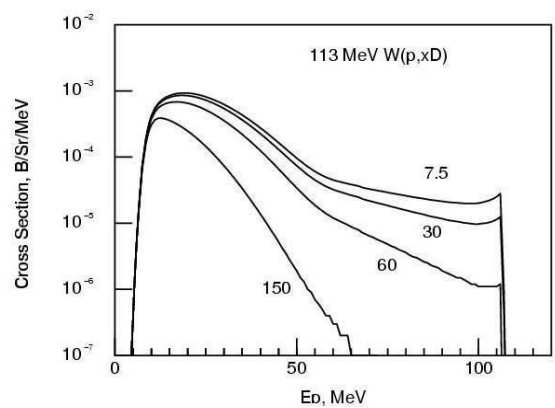

Fig.10. Double differential cross sections of deuteron emission in p+$^{184}$W reactions are calculated at the energy 113.0MeV.

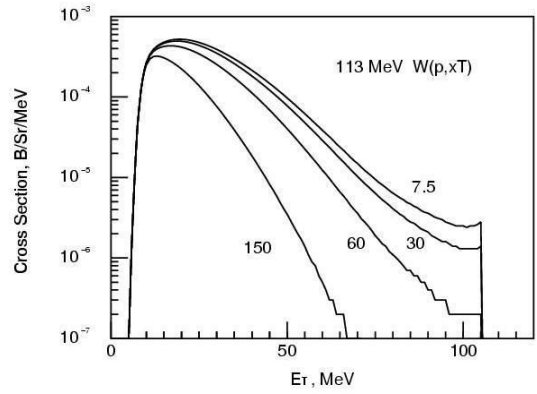

Fig.11. Calculated double differential cross sections of triton emission in p+$^{184}$W reactions are given at the energy 113.0MeV.

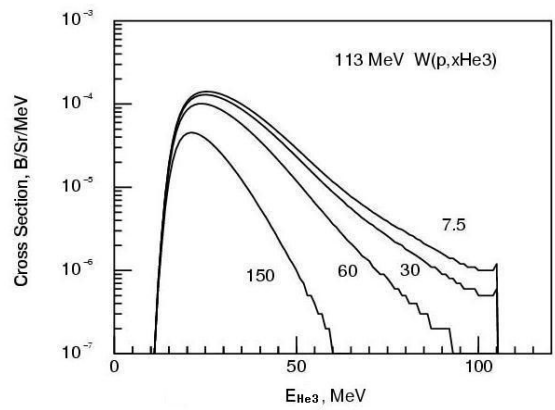

Fig.12. Double differential cross sections of helium-3 emission in p+$^{184}$W reactions are obtained at the energy 113.0MeV.

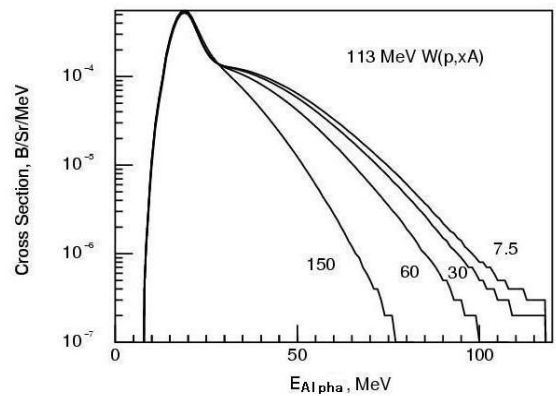

Fig.13. Double differential cross sections of alpha-particle emission in p+$^{184}$W reactions are calculated at the energy 113.0MeV.

# 4   Conclusions

A set of suitable proton optical potential parameters are obtained at incident proton energy up to 250MeV based on the experimental data of non-elastic cross-sections and elastic scattering angular distributions for p+$^{184}$W reactions. The theoretical models including optical model, intra-nuclear cascade model, distorted born wave approximation theory, pre-equilibrium and equilibrium reaction theories provide the good description of the shapes and magnitude of the energy spectra and double differential cross sections of proton, neutron, deuteron, triton, helium and alpha particle emissions. The calculated results are useful for design of ADS and other equipments.